\documentclass[prb,twocolumn,amsmath,amssymb,superscriptaddress,showpacs]{revtex4-1}

\usepackage{bm}
\usepackage{times}
\usepackage{amsmath}
\usepackage{mathrsfs}
\usepackage{graphicx}
\usepackage{epstopdf}
\usepackage[colorlinks=true, letterpaper=ture, pdfstartview=FitV, linkcolor=blue, citecolor=blue, urlcolor=blue]{hyperref}

\begin{document}
	
\title{Spatial spin-spin correlations of the single-impurity Anderson model with a ferromagnetic bath}
\author{Peng Fan}
\affiliation{Theoretical Condensed Matter Physics and Computational Materials Physics Laboratory, College of Physical Sciences, University of Chinese Academy of Sciences, Beijing 100049, China}
\affiliation{School of Electronic, Electrical and Communication Engineering, University of Chinese Academy of Sciences, Beijing 100049, China}
\author{Ning-Hua Tong}
\email{nhtong@ruc.edu.cn}
\affiliation{Department of Physics, Renmin University of China, 100872 Beijing, China}
\author{Zhen-Gang Zhu}
\email{zgzhu@ucas.ac.cn}
\affiliation{Theoretical Condensed Matter Physics and Computational Materials Physics Laboratory, College of Physical Sciences, University of Chinese Academy of Sciences, Beijing 100049, China}
\affiliation{School of Electronic, Electrical and Communication Engineering, University of Chinese Academy of Sciences, Beijing 100049, China}
\affiliation{CAS Center for Excellence in Topological Quantum Computation, University of Chinese Academy of Sciences, Beijing 100190, China}
\date{\today}
\begin{abstract}
  We investigate the interplay between the Kondo effect and the ferromagnetism by an one dimension Anderson impurity model with a spin partially polarized bath, using the projective truncation approximation under Lacroix basis.
  The equal-time spatial spin-spin correlation function (SSCF) is calculated. For the case of spin-unpolarized conduction electrons, it agrees qualitatively with the results from density matrix renormalization group (DMRG).
  For system with partially spin-polarized conduction electrons, an oscillation in the envelope of SSCF emerges due to the beating of two Friedel oscillations associated to two spin-split Fermi surfaces of conduction electrons. The period is proportional to the inverse of magnetic field $h$.  A fitting formula is proposed to perfectly fits the numerical results of SSCF in both the short- and long-range regions.
  For large enough bath spin polarization, a bump appears in the curve of the integrated SSCF. It marks the boundary between the suppressed Kondo cloud and the polarized bath sites.

\end{abstract}
\maketitle

\section{Introduction}
Kondo effect, a well-known many-body strong correlation phenomena arising from the exchange coupling between the local spin and that of the conduction electrons, has been investigated for a long time~\cite{Hewson1993, Mitchell2011}.
Most of the previous investigations focus on the case where the local spin is coupled to paramagnetic electrodes.~\cite{GoldhaberGordon1998, Gruner1974}
However, in order to investigate the interplay between the Kondo effect and the ferromagnetism, the systems with a quantum dot (QD) coupled to ferromagnetic electrodes are intensively studied in theories ~\cite{Zhang2002,Martinek2003, Martinek2003a, Choi2004, Martinek2005, Sindel2007, Gaass2011} and experiments ~\cite{Pasupathy2004, Hauptmann2008, Hofstetter2010}.

Theoretical results for such a system, in particular, those of the numerical renormalization group (NRG) method, show that the Kondo resonance at the Fermi surface is suppressed and splits into two peaks. This is attributed to the QD level splitting, resulting from the spin dependent level renormalization induced by the charge fluctuation between the QD and the electrodes ~\cite{Martinek2003, Martinek2003a}.
An effective exchange field $\bm{B}_{\text{ex}}$ characterizes the level splitting.
It is observed that if an external magnetic field $\bm{B}=-\bm{B}_{\text{ex}}$ is applied to the QD, the Zeeman energy will compensate the level splitting and restore the Kondo resonance ~\cite{Martinek2003a}.
According to Haldane's scaling method ~\cite{Haldane1978},
there are two kinds of charge fluctuation processes contributing to the level splitting, the electron-like  process and the hole-like process, which can compensate each other by properly tuning the QD level energy $\varepsilon_{\text{d}}$ (or the gate voltage $V_{\text{g}}$) in the absence of $\bm{B}$, resulting in the disappearance of the Kondo resonance splitting ~\cite{Martinek2005, Sindel2007}.
The restored Kondo resonance occurs either in the local-moment regime or in the mixed-valence regime, strongly relying on the shape of the density of state (DOS) of the conduction electrons ~\cite{Martinek2005}.
As $\varepsilon_{\text{d}}$ approaches the charge resonance regime, the QD level splitting logarithmically diverges as $\ln\left(|\varepsilon_{\text{d}}|/|U+\varepsilon_{\text{d}}|\right)$ for a flat conduction band, which indicates that the spin splitting results from many-body correlation effects ~\cite{Martinek2005, Sindel2007}.
The $\varepsilon_{\text{d}}$ ($V_{\text{g}}$) dependence of the level splitting also provides an applicable way to tune the local spin direction, which is of importance in the spintronics ~\cite{Martinek2005, Sindel2007, Hauptmann2008}.
These theoretical results have already been confirmed by the measurement of the differential conductance in QD systems, such as C$_{60}$ molecules~\cite{Pasupathy2004} or carbon-nanotube (CNT)~\cite{Hauptmann2008} coupled to ferromagnetic nickel electrodes, CNT coupled to ferromagnetic PdNi elctrondes,~\cite{Gaass2011} and the superconductor-QD-ferromagnet hybrid device ~\cite{Hofstetter2010}.

These works mainly explored how the local properties are influenced by the spin polarized electrodes, studying quantities like the QD spectral function, conductance, differential conductance, or the occupation of the QD level~\cite{Zhang2002,Bulka2003,Choi2004,Gaass2011,Martinek2003,Martinek2003a,Martinek2005, Sindel2007,Wrzesniewski2019,Liu2019}. Compared to the local properties, far less is known on how the non-local properties are influenced by the ferromagnetic electrodes. Here we consider the equal-time spatial SSCF between the spins of QD and conduction electrons. It can give a snapshot for the profile of Kondo screening cloud. Considering the
difficulty and controversy in the experimental observation of Kondo screening cloud due to its large spatial scale~\cite{Affleck2001,Park_2013,Madhavan1998,Manoharan2000,Prueser2011,Borzenets2020}, the study of influence of ferromagnetic electrodes on SSCF may provide useful information that facilitates experimental observation of Kondo screening cloud, considering that the size of Kondo cloud could be suppressed by the spin polarization of electrodes (see below).

For the QD system with paramagnetic electrodes, the non-local properties have been well studied~\cite{Sorensen1996,Barzykin1998,Borda2007,Holzner2009}.
In the paramagnetic case, it is well known that the localized spin on QD is screened by the surrounding conduction electrons when the temperature goes below the Kondo temperature $T_{\text{K}}$. At zero temperature, it is completely screened out and the ground state becomes a Kondo singlet.
In Norzi\'eres Fermi liquid theory~\cite{Nozieres1974}, the conduction electrons form a screening cloud and spread in a spatial region.
Scaling theory shows that the range of the screen cloud is $\xi_{\text{K}}= \hbar v_{\text{F}}/T_{\text{K}}$ ($v_{\text{F}}$ is the Fermi velocity)~\cite{Barzykin1998, Ribeiro2019}. %
According to NRG calculation, at zero temperature the equal-time spatial SSCF crosses over from $x^{-1}$ to $x^{-2}$ around $\xi_{\text{K}}$, where $x$ is the distance from the impurity~\cite{Barzykin1998,Hand2006,Borda2007}.
At finite temperature, another spatial scale $\xi_{T}$ emerges, which cuts off the long-range power law behavior replaced by an exponential behavior~\cite{Barzykin1998,Borda2007}.

In this paper, we attempt to explore how the partially spin-polarized conduction bath (ferromagnetic electrodes) influences the equal-time spatial SSCF at zero temperature.
We study the Anderson's impurity model (AIM) with a spin polarized conduction bath using the equation of motion of two-time Green's functions with projective truncation approximation (PTA-EOMGF)\cite{Fan2018,Fan2019}. This method is a many-body numerical calculation technique recently proposed by two of the authors of the paper. We found that the effect of bath spin polarization on SSCF is a two fold. First, an additional oscillation with period proportional to $1/h$ emerges in the amplitude of SSCF (Here $h$ is the effective Zeeman field in the spin polarized electrodes), due to the beating of Friedel oscillations associated to two spin-split Fermi surfaces of the bath electrons. Second, the bath spin polarization induces a spatial scale $x_b \sim 1/h$ in the integrated SSCF $\Sigma(x)$, at which the exponential decay in the intermediate $x \ll x_b$ crosses over to linear decay in $x \gg x_b$ regime where the conduction electrons are spin polarised. For sufficiently large $h$, i.e., $h \gg T_K$, $x_b$ marks the crossover from the compressed Kondo screening cloud to spin polarized bath sites. This shows that the size of the Kondo cloud can be suppressed by a sufficiently large bath spin polarization, but the scaling behavior is preserved in the short-range region. This may be exploited to design an experimental setup for studying the Kondo cloud.

The rest of the paper is organized as follows. In Sec. II, we introduce the model Hamiltonian and define all the quantities appearing in the paper.
In Sec. III, we introduce the theoretical aspect of the method.
In Sec. IV, we show the numerical results and discuss the influence of the spin polarization on the Kondo effect.
In Sec. V, we summarize this paper.

\section{Model and Definitions}

\begin{figure}[t!]
	\begin{center}
		\includegraphics[width=0.9\columnwidth]{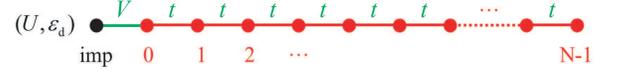}
	\end{center}
	\caption{(Color online) The AIM studied in this work. The black and red circles represent the impurity and the semi-infinite conduction electron chain, respectively.} \label{Fig1}
\end{figure}

The Anderson models which describe a QD coupled to ferromagnetic electrodes are not unique~\cite{Martinek2003,Martinek2003a,Martinek2005,Simon2007}. Here, we consider an impurity coupled to a semi-infinite conduction electron chain (see Fig. \ref{Fig1}). The spin of the conduction electrons is partially polarized by an effective magnetic field $h$, which lifts the spin degeneracy of the conduction electrons. Here, $h$ arises from the mean-field description of the exchange interaction between conduction electrons. The spin asymmetry in the electrodes is therefore described by the spin dependent density of states. The Hamiltonian of AIM studied in this work is given by~\cite{Gazza2006}
\begin{equation} \label{Hamilton}
   H = H_{\text{bath}} + H_{\text{imp}} + H_{\text{hyb}},
\end{equation}
where
\begin{align} \label{comp_Hamilton}
   H_{\text{bath}} &=  - t \sum_{i=0,\sigma}^{N-2} \left( a^{\dagger}_{i\sigma} a_{i+1\sigma} + \it{h.c.} \right)  -
   \sum_{i=0,\sigma}^{N-1} \mu a^{\dagger}_{i\sigma} a_{i\sigma} \nonumber\\
   &\,\,\,\,\,\,\, + \sum_{i=0}^{N-1} h s^{z}_{i},\\
   H_{\text{hyb}} &= V \sum_{\sigma} \left(a^{\dagger}_{0\sigma} d_{\sigma} + d^{\dagger}_{\sigma} a_{0\sigma}\right),\\
   H_{\text{imp}} &= U n_{\text{d} \uparrow} n_{\text{d} \downarrow} + \sum_{\sigma} \left(\varepsilon_{\text{d}} - \mu\right) d^{\dagger}_{\sigma} d_{\sigma}.
\end{align}
%
$a^{\dagger}_{i\sigma} (a_{i\sigma})$ is the creation (annihilation) operator of the conduction electron, $d^{\dagger}_{\sigma} (d_{\sigma})$ is the impurity creation (annihilation) operator. $t = 1$ is the nearest neighbor hopping, which is set as the energy unit. $\mu$ is the chemical potential and is set zero throughout this work. $h$ is the external magnetic field. $s^{z}_{i}= (n_{i\uparrow} - n_{i\downarrow})/2$. $N$ is the number of bath sites. $V$ is the strength of the hybridization between the impurity and the first site of the chain. $U$ is the Coulomb repulsion energy and $\varepsilon_{\text{d}}$ is the on-site energy of the impurity. This Hamiltonian, after proper rescaling of parameters, has identical structure of conduction electron band as in the work of Holzner~\cite{Holzner2009}.
The degree of the spin polarization is defined by
\begin{equation} \label{Rpol}
R_{\text{pol}}(h)=\frac{\sum_{i = 0}^{N-1} \left(\langle n_{i\uparrow} \rangle - \langle n_{i\downarrow} \rangle \right)}{\sum_{i = 0}^{N-1} \left( \langle n_{i\uparrow} \rangle + \langle n_{i\downarrow} \rangle \right)},
\end{equation}
where $-1\le R_{\text{pol}}(h)\le +1$. The sign of $R_{\text{pol}}(h)$ denotes the direction of the spin polarization in $z$ axis. The equal-time spatial SSCF between the impurity and the conduction electron at position $i$ $(i=0,1,\cdots,N-1)$ is given by
\begin{equation}
\chi(i) = \langle \vec{s}_{i} \cdot \vec{s}_{\text{d}} \rangle,
\end{equation}
which snapshots the spatial extension of the screening cloud. In this paper, we merely consider the zero temperature case. Therefore, $\langle \cdots \rangle$ denotes the ground state average. Because of the external magnetic field, SU(2) symmetry of the model is broken in $z$-direction. In $xy$-plane, the spin is still isotropic. Therefore, we divide $\chi(i)$ into the transverse and longitudinal parts
\begin{equation} \label{chi}
\chi(i) = \chi_{xy}(i) + \chi_{z}(i).
\end{equation}
Here,
\begin{equation} \label{chixy}
\chi_{xy}(i) =
\langle a_{i\uparrow}^{\dag} a_{i\downarrow} d_{\downarrow}^{\dag} d_{\uparrow} \rangle
\end{equation}
denotes the spin-spin correlation between the $x$ and $y$ components of the spin.
\begin{equation} \label{chiz}
\chi_{z}(i) =
\frac{1}{4} \sum_{\sigma} \left(\langle n_{i\sigma} n_{d\sigma} \rangle - \langle n_{i\sigma}n_{d\bar{\sigma}} \rangle\right),
\end{equation}
is the correlation function between the $z$ component of the spins. $\bar{\sigma}$ represents the opposite spin of $\sigma$.
Detailed expressions of $\chi_{xy}(i)$ and $\chi_{z}(i)$ are given in Appendix A. We also define the integrated equal-time spatial SSCF
\begin{equation}  \label{sigma}
\Sigma(x) = 1 + \sum_{i=0}^{x} \frac{\langle \vec{s}_{i}\cdot \vec{s}_{d} \rangle}{\langle s_{d}^{2} \rangle},
\end{equation}
which is the same as Holzner's definition.~\cite{Holzner2009} $\Sigma(x)$ describes the extent to which the local spin at the impurity site is screened by the conduction electrons from the position $i=0$ to $i=x$. When the local spin is fully screened, $\Sigma(N-1) = 0$ for a sufficiently long chain, due to the formation of the Kondo singlet. As for the case where the spin of the conduction electron is partially polarized by the external magnetic field, the bath polarization can induce a partial polarization of the impurity spin in the opposite direction, which leads to $ \Sigma(N-1) < 0$ (see Fig. \ref{Fig7}). In this case the impurity spin is partially screened.

\section{Numerical calculation }

The calculation of the equal-time spatial SSCF is challenging for most of methods used to investigate AIM, such as perturbative method and NRG approach.
Recently, two of the authors of the present work proposed a many-body calculation method, PTA-EOMGF~\cite{Fan2018,Fan2019} and applied it to Anderson impurity model.
In that work, the continuous bath degrees of freedom is discretized using the NRG discretization formula, which improves the energy resolution at the cost of losing the spatial resolution.~\cite{Wilson1975}
In this work, in order to improve the spatial resolution, instead of using the NRG discretization formula, we directly diagonalize $H_{\text{bath}}$ using a unitary transformation. This trick balances the energy resolution and the spatial resolution.
In matrix form, $H_{\text{bath}}$ is written as
\begin{equation} \label{Hbath_mat_form}
H_{\text{bath}} = \sum_{\sigma} \bm{a}_{\sigma}^{\dagger} \bm{H}_{\sigma} \bm{a}_{\sigma} - \mu\sum_{\sigma} \bm{a}_{\sigma}^{\dagger} \bm{a}_{\sigma},
\end{equation}
where

\begin{equation} \label{Hbath_mat}
\bm{H}_{\sigma} = -\left(
\begin{array}{cccc}
 - \frac{1}{2}h\sigma   & t                              & 0                                        &\cdots \\
t                              & - \frac{1}{2}h \sigma   & t                              &\cdots \\
0                                        & t                                & - \frac{1}{2}h \sigma   &\cdots \\
\vdots                               & \vdots                                & \vdots                               &
\end{array}
\right),
\end{equation}
and $\bm{a_{\sigma}} = \left( a_{0\sigma}, a_{1\sigma}, \cdots, a_{N-1\sigma} \right)^{\mathcal{T}}$.
$\sigma = 1$ ($\sigma = -1$ ) for spin up (down).
The superscript $\mathcal{T}$ denotes the matrix transpose.
In the following, we always use a bold symbol to denote a matrix.
$\bm{H}_{\sigma}$ is an Hermitian matrix, which is diagonalized by a unitary matrix
\begin{equation} \label{d_Hbath_mat}
\bm{H}_{\sigma} = \bm{U}_{\sigma} \bm{\Lambda}_{\sigma} \bm{U}_{\sigma}^{\mathcal{T}}.
\end{equation}
Here $\bm{\Lambda}_{\sigma}$ is a diagonal matrix and $\left( \bm{\Lambda}_{\sigma} \right)_{kk} = \varepsilon_{k\sigma}$.
$\bm{U}_{\sigma}$ is a real unitary matrix. We have $\bm{U}^{\mathcal{T}}_{\sigma}\bm{U}_{\sigma} = \bm{U}_{\sigma}\bm{U}^{\mathcal{T}}_{\sigma} = \bm{1}$, where $\bm{1}$ is the identity matrix. Substituting Eq. (\ref{d_Hbath_mat}) into Eq. (\ref{Hbath_mat_form}), we get
\begin{equation}
\mathcal{H}_{\text{bath}} = \sum_{\sigma} \bm{c}_{\sigma}^{\dagger} \bm{\Lambda}_{\sigma} \bm{c}_{\sigma} - \mu \sum_{\sigma} \bm{c}_{\sigma}^{\dagger} \bm{c}_{\sigma},
\end{equation}
where
\begin{equation}
\bm{c}_{\sigma} = \bm{U}_{\sigma}^{T} \bm{a}_{\sigma},
\,\,\,\,\,\,\,\,\,\
\bm{c}_{\sigma}^{\dagger} = \bm{a}_{\sigma}^{\dagger} \bm{U}_{\sigma}.
\end{equation}
The inverse transform is readily obtained
\begin{equation} \label{inv_trans}
\bm{a}_{j\sigma} = \sum_{k}\bm{U}_{jk\sigma} \bm{c}_{k\sigma},
\,\,\,\,\,\,\,\,\,\
\bm{a}_{j\sigma}^{\dagger} = \sum_{k}\bm{c}_{k\sigma}^{\dagger} \bm{U}_{jk\sigma}.
\end{equation}
For $H_{\text{hyb}}$, Eq. (\ref{inv_trans}) gives us
\begin{equation}
\mathcal{H}_{\text{hyb}} = \sum_{k\sigma} V_{k\sigma} \left(c_{k\sigma}^{\dagger}d_{\sigma} + d_{\sigma}^{\dagger}c_{k\sigma}\right),
\end{equation}
where $V_{k\sigma} = V U_{0k\sigma}$.
Consequently, we obtain the Hamiltonian $\mathcal{H}$ in the diagonal representation of $H_{\text{bath}}$
\begin{align} \label{Hk}
\mathcal{H} &= \sum_{k\sigma} \left( \varepsilon_{k\sigma} - \mu \right) c_{k\sigma}^{\dagger} c_{k\sigma} + \sum_{k\sigma} V_{k\sigma} \left(c_{k\sigma}^{\dagger}d_{\sigma} + d_{\sigma}^{\dagger}c_{k\sigma}\right) \nonumber \\
&\,\,\,\,\,\,\,+ U n_{d\uparrow}n_{d\downarrow} + \sum_{\sigma} \left( \varepsilon_{d} - \mu \right) d_{\sigma}^{\dag}d_{\sigma}.
\end{align}
Substituting Eq. (\ref{inv_trans}) into Eq. (\ref{chixy}) and Eq. (\ref{chiz}), we obtain
\begin{equation} \label{chixy_k_to_i}
\chi_{xy}(i) = \sum_{k_{1}k_{2}} \bm{U}_{ik_{1}\uparrow}\bm{U}_{ik_{2}\downarrow} \langle c_{k_{1}\uparrow}^{\bm{\dag}} c_{k_{2}\downarrow} d_{\downarrow}^{\dag} d_{\uparrow} \rangle
\end{equation}
and
\begin{equation} \label{chiz_k_to_i}
	\chi_{z}(i) = \frac{1}{4} \sum_{k_{1}k_{2}\sigma} \bm{U}_{ik_{1}\sigma} \bm{U}_{ik_{2}\sigma} \langle c_{k_{1}\sigma}^{\dagger} c_{k_{2}\sigma} \left( n_{d\sigma} - n_{d\bar{\sigma} } \right) \rangle.
\end{equation}
In the PTA-EOMGF approach, $\chi_{xy}(i)$ and $\chi_{z}(i)$ are evaluated as full two-body correlation functions. The same quantities have also been studied by the finite-U slave boson mean-field approximation method~\cite{Ribeiro2019,Buesser2010}.
In the same way, the spin dependent occupancy at the position $i$ is expressed as
\begin{equation} \label{n_k}
	\langle n_{i\sigma} \rangle = \sum_{k_{1}k_{2}} U_{ik_{1}\sigma} U_{ik_{2}\sigma} \langle c_{k_{1}\sigma}^{\dagger} c_{k_{2}\sigma} \rangle.
\end{equation}

The averages appearing in Eq. (\ref{chixy_k_to_i}), Eq. (\ref{chiz_k_to_i}) and Eq. (\ref{n_k})
are directly calculated by
PTA-EOMGF.~\cite{Fan2018,Fan2019}
We take the Lacroix basis $\vec{A} = \{A_{1}, A_{2k} , A_{3}, A_{4k}, A_{5k}, A_{6k}\},( k = 1, 2, \dots, N)$, where~\cite{Lacroix1981}
\begin{alignat}{3}
A_{1} &= d_{\sigma},\,\,\,&A_{2k}& = c_{k\sigma},\,\,\, &A_{3}&= n_{\bar{\sigma}}d_{\sigma} \nonumber\\
A_{4k} &= n_{\bar{\sigma}}c_{k\sigma},\,\,\,&A_{5k}& = d^{\dagger}_{\bar{\sigma}}c_{k\bar{\sigma}}d_{\sigma},\,\,\,&A_{6k}&= c^{\dagger}_{k\bar{\sigma}}d_{\bar{\sigma}}d_{\sigma}.
\end{alignat}
The equation of motion of the retarded Green's function matrix is
\begin{equation} \label{EOM_GF}
\omega \bm{G} (\vec{A} | \vec{A}^{\dagger} )_{\omega} = \langle \{ \vec{A}, \vec{A}^{\dagger} \} \rangle + \bm{G}( [ \vec{A}, \mathcal{H} ] | \vec{A}^{\dagger} )_{\omega},
\end{equation}
where $\omega$ is the frequency. In general, the component of the commutator $[\vec{A}, \mathcal{H}]$ reads
\begin{equation} \label{commutator_A_and_H}
[A_{i},\mathcal{H}] = \sum_{j=1} \bm{M}_{ji} A_{j} + B_{i},
\end{equation}
where $B_{i}$ is a new higher order operators outside the basis $\vec{A}$.
We define the inner product of two arbitrary operators $A$ and $B$ as
\begin{equation}
(A|B) \equiv \langle \{ A^{\dagger}, B\} \rangle.
\end{equation}
The curly bracket denotes the anticommutator. $\langle \hat{O}\rangle = \text{Tr}(\rho \hat{O})$, where $\rho = e^{-\beta \mathcal{H}}/\text{Tr}(e^{-\beta \mathcal{H}})$ is the equilibrium density operator and $\beta = 1/T$ ($T$ is the temperature). In this paper, we take the nature unit. In order to truncate the EOMGF, we project Eq. (\ref{commutator_A_and_H}) to the basis operator $A_{k}$,
\begin{equation}
\bm{L} = \bm{I}\bm{M}+\bm{P},
\end{equation}
where $\bm{L}_{ki} = (A_{k}|[A_{i},\mathcal{H}])$, $\bm{I}_{kj} = (A_{k}|A_{j})$, and $\bm{P}_{ki}=(A_{k}|B_{i})$. We neglect those components of $B_{i}$ that are orthogonal to the basis set, i.e., $B_{i} \approx \sum_{j=1} \bm{N}_{ji} A_{j}$. Therefore, $\bm{P} = \bm{I}\bm{N}$. By defining $\bm{M}_{\text{t}} = \bm{M} + \bm{N}$, we have $\bm{L} = \bm{I}\bm{M}_{\text{t}}$. Hence,
\begin{equation} \label{commutator_A_and_H_apprx}
[\vec{A}, \mathcal{H}] \approx \bm{M}^{\mathcal{T}}_{\text{t}} \vec{A}.
\end{equation}
Substituting Eq. (\ref{commutator_A_and_H_apprx}) into Eq. (\ref{EOM_GF}), we obtain a formal approximate solution
\begin{equation}
G(\vec{A}|\vec{A}^{\dagger})_{\omega} \approx (\omega \bm{1} - \bm{M}^{\mathcal{T}}_{\text{t}})^{-1} \bm{I}^{\mathcal{T}}.
\end{equation}
We calculate $G(\vec{A}|\vec{A}^{\dagger})_{\omega}$ by solving a generalized eigenvalue problem numerically. More details about PTA-EOMGF can be found in Refs.~\onlinecite{Fan2018,Fan2019}.
The averages are directly calculated by the corresponding retarded Green's function via the fluctuation-dissipation theorem
\begin{equation}
\langle A_{i}A_{j} \rangle = -\frac{1}{\pi}\int^{+\infty}_{-\infty} d\omega \frac{  \Im\left[ G(A_{j}|A_{i})_{\omega} \right] }{e^{\beta\omega} + 1},
\end{equation}
where $\Im\left[ G(A_{j}|A_{i})_{\omega} \right]$ denotes  the imaginary part of $G(A_{j}|A_{i})_{\omega} $.

The unitary transformation defines a correspondence between the real space and the energy space. The chain form of Hamiltonian used in this work is equivalent to a uniform linear discretization in energy space. As a result, the spatial correlation between any two positions is obtained at the cost of reducing the energy resolution. If we want to improve the energy resolution, we need to increase the length of the conduction electron chain. In this work, we typically used chain length $N = 500 \sim 1000$, which is sufficient for discussing the competition between Kondo screening and bath spin polarization.

\section{Results and Discussions}

In this section, we will show the results of two cases: paramagnetic case and the ferromagnetic case. For the paramagnetic case, the conduction electron bath is in the absence of the external magnetic field and $R_{\text{pol}}= 0$. For the ferromagnetic case, an effective magnetic field $h$ is applied to the conduction electron bath to mimic the internal field from the ferromagnetic exchange interaction. It induces $R_{\text{pol}}\neq 0$. Here, we first give the results of the paramagnetic case in order to benchmark our numerical calculation. For the ferromagnetic case, we discuss the competition between Kondo screening and bath spin polarization in terms of the equal-time spatial SSCF.

\subsection{Paramagnetic case: $R_{\text{pol}} = 0$}
\begin{figure}[t!]
	\begin{center} \includegraphics[width=1.0\columnwidth]{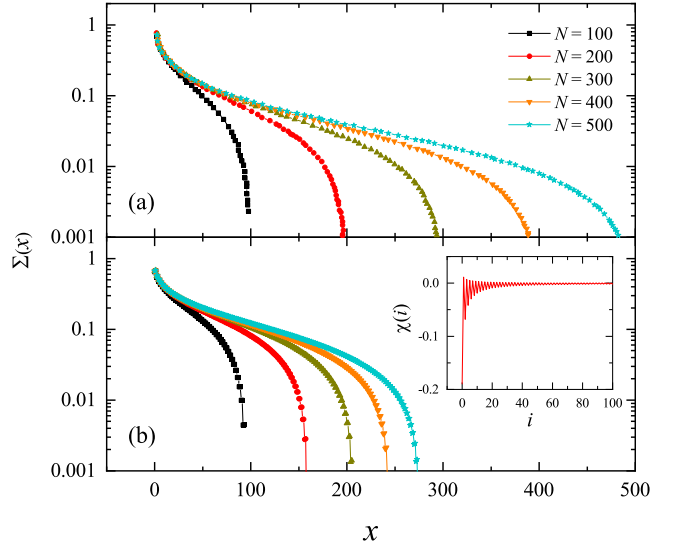}
	\end{center}
	\caption{ (Color online) The dependence of $\Sigma(x)$ on the chain index $x$ for various conduction electron chains ($N = 100$, $200$, $300$, $400$, $500$) at $T=0.0$, $U=1.0$, $\varepsilon_{d} = -U/2$, $V = \sqrt{0.2}$, and $h = 0.0$. This set of parameters is equivalent to that used in Fig. 1 of Holzner's work~\cite{Holzner2009}. (a) shows the result of DMRG, which is obtained by digitizing Fig. 1 in Ref.\cite{Holzner2009}. (b) the result of PTA-EOMGF. The inset shows the dependence of $\chi(i)$ on the chain index $i$  for a chain of length $N=300$.  } \label{Fig2}
\end{figure}
Following Eq. (\ref{chi}), Eq. (\ref{chixy_k_to_i}), and Eq. (\ref{chiz_k_to_i}), we use the extended PTA-EOMGF method to calculate $\chi(i)$.
 The inset of Fig. \ref{Fig2}(b) shows the dependence of $\chi(i)$ on the site index $i$. $\chi(i)$ has an even-odd oscillation between positive and negative values. The magnitude decays algebraically. It has been known that the decay of the spin-spin correlation with distance $i$ crosses over from $i^{-1}$ to $i^{-2}$ around $\xi_{\text{K}}$~\cite{Borda2007, Barzykin1998}. Our result is in qualitative agreement with both the DMRG result and the perturbation result~\cite{Sorensen1996, Holzner2009}.

We further sum up $\chi(i)$ according to Eq. (\ref{sigma}) and show the obtained $\Sigma(x)$ in the main panel of Fig. \ref{Fig2}(b).
 To benchmark the PTA-EOMGF method, we compare our result with that of DMRG (see Fig. \ref{Fig2}(a)).
 The DMRG result is obtained by digitizing Fig. 1 in  Holzner's work~\cite{Holzner2009}.
Here, we take five various conduction electron chains ($N = 100, 200, 300, 400, 500$) in order to show the finite-sized effect. Both results share some common features. For each $N$, $\Sigma(x)$ decays quickly at small $x$, then transits into an exponential decay before it finally drops to zero at some point. With increasing $N$, the intermediate range with stabled exponential decay enlarges, showing that the the finite-size effect is reduced gradually. For the present Lacroix basis, $\Sigma(x)$ calculated via PTA-EOMGF decays with $x$ faster than that of DMRG, indicating that the Kondo screening length scale $\xi_{\text{K}}$ obtained by PTA-EOMGF is smaller than that of DMRG.

 The qualitative correctness of PTA results is not surprising. The original Lacroix approximation already well describes the Kondo effect because the operators of spin exchange have been kept in the truncation approximation. Our PTA further optimizes the truncation using operator projection. Quantitatively, a tiny breaking of SU(2) symmetry, i.e., a $10^{-4}$ relative difference between SSCFs in $z$ and $x$ ($y$) directions, is observed in PTA result. This does not influence $\Sigma(x)$ in small $x$ regime (corresponding to high energy part), but makes $\Sigma(x)$ in large $x$ regime (low energy part) less accurate. In particular, it accelerates the decay of $\Sigma(x)$ and makes it negative in the large $x$ regime (not shown in Fig.2(b)). As a result, the exponential behavior of $\Sigma(x)$ in the intermediate $x$ regime is sustained in a smaller region of $x$ and the obtained $\xi_K$ is also smaller in comparison with DMRG result.

\begin{figure}[t!]
	\begin{center}
		\includegraphics[width=1.0\columnwidth]{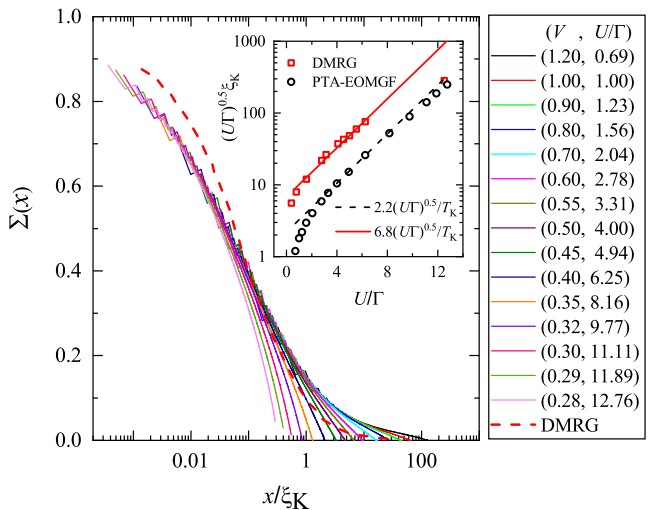}
	\end{center}
	\caption{ (Color online) The scaling analysis of $\Sigma(x)$. The dash line shows the scaling function obtained by DMRG. By scaling analysis, we obtain the screen length scale $\xi_{\text{K}}$, which is shown in the inset (black cycles). The red squares are the result of DMRG. We take $T=0.0$, $U=1.0$, $\varepsilon_{d} = -U/2$, $h = 0.0$, and $N=801$. DMRG data are obtained by digitizing Fig. 2(b) in Holzner's work~\cite{Holzner2009}.} \label{Fig3}
\end{figure}

According to the discussion of Holzner \emph{et. al.},  $\Sigma(x)$ has a universal form characterized by the Kondo screening length $\xi_{\text{K}}$, when $\xi_{\text{K}}$ is significantly shorter than the length of the conduction electron chain $N$~\cite{Holzner2009}.
Theoretically, $\xi_{\text{K}}$ is determined by the Kondo temperature $T_{\text{K}}$ via $\xi_{\text{K}} = \hbar v_{\text{F}}/ T_{\text{K}}$. It  depends on $U$, $V$ and $\varepsilon_{d}$ in the form ~\cite{Holzner2009}
\begin{equation} \label{xi_k}
\xi_{\text{K}} = \frac{\hbar v_{\text{F}}}{\sqrt{U\Gamma}} \exp\left[ \frac{\pi|\varepsilon_{d}| |\varepsilon_{d}+U|}{2U\Gamma}\right],
\end{equation}
where $\Gamma = V^{2}/t$. Note that for both Holtzner's and our Hamiltonians, $\hbar v_{\text{F}}=2.0$ in the natural unit.
To further benchmark our numerical method, we do a scaling analysis of $\Sigma(x)$ to extract the screening length $\xi_{\text{K}}$ and  compared it with DMRG result.
 Due to the finite-sized effect, the numerical result of  $\Sigma(x)$  deviates from the universal form gradually with increasing $x$ (see Fig. \ref{Fig2}).
Eq. (\ref{xi_k}) indicates that at the condition $\varepsilon_{d} = -U/2$, $\xi_{\text{K}}$ reduces gradually with the reduce of $U/\Gamma$. Thus for smaller $U/\Gamma$, the condition $\xi_{\text{K}}\ll N$ is met better and the finite-sized effect is smaller.
Therefore, in the scaling  process, we pin the curve with the smallest $U/\Gamma$, namely the curve of $V=1.2$, and collapse other curves from the smallest to the largest $U/\Gamma$ onto the curve of $V=1.2$ by adjusting $\xi_{\text{K}}$.
The inset of Fig. \ref{Fig3} shows the dependence of $(U\Gamma)^{0.5}\xi_{\text{K}}$ on $U/\Gamma$  on a semilog plot, where $\xi_{\text{K}}$ is extracted from the scaling analysis of $\Sigma(x)$.
We use $p(U\Gamma)^{0.5}/T_{\text{K}}$ to fit our result. It is found that our result is fitted well by the function $p(U\Gamma)^{0.5}/T_{\text{K}}$, obtaining $p=2.2$, which suggests that $\xi_{\text{K}}$ obtained by PTA-EOMGF is consistent with Eq. (\ref{xi_k}).
The red squares are the result of DMRG, which are fitted as well. Holzner \emph{et. al.} obtained $p=6.8$, which is $3.1$ times larger than our result~\cite{Holzner2009}.
To compare the universal function $\Sigma(x/\xi_{\text{K}})$ of PTA-EOMGF with that of DMRG, we use $3.1$ times of our $\xi_{\text{K}}$ in the plot (see the main panel of Fig. \ref{Fig3}). It is found that both methods produce similar $\Sigma(x/\xi_{\text{K}})$ curve. PTA-EOMGF result decreases slower with increasing $x/\xi_{\text{K}}$.

In the inset of Fig. (\ref{Fig3}), we find that $\xi_{\text{K}}$ obtained by PTA follows Eq. (\ref{xi_k}) in a larger region than that of DMRG. Due to finite size effect, a perfect scaling analysis is only possible when $\xi_{\text{K}}$ is much smaller than the length of chain $N$. Since DMRG produces larger $\xi_{\text{K}}$, i.e., larger Kondo cloud, the finite length of the chain exerts a greater influence on the scaling analysis and makes $\xi_{\text{K}}$ deviate early from the theoretical curve as $U/\Gamma$ increases. In contrast, PTA gives a smaller $\xi_{\text{K}}$ for the same $U/\Gamma$, the finite-size effect has a smaller influence on the scaling analysis. The scaling analysis can thus be preformed well for larger $U/\Gamma$ for the same chain length $N$.

It should be noted that in Fig.3, the data collapse is done in the linear-log plot. By using this plot, we effectively increase the weight of data in the small $x$ regime in the fitting, which is more accurate than the data in large $x$ regime. The exponential form in $\Sigma(x)$ in the intermediate $x$ regime (see Fig.2) calls for a data collapse analysis in the log-linear plot. However, since the data quality is poorer in the large $x$ (i.e., small $\Sigma(x)$) regime due to finite size effect and $SU(2)$ symmetry breaking error, we find it much difficult to perform the analysis in log-linear plot.

To conclude, the equal-time spatial SSCF obtained by PTA-EOMGF is not quantitatively comparable with that of DMRG, but are qualitatively correct. It correctly follows the theoretical expectation (see Eq. (\ref{xi_k})).
The precision of PTA-EOMGF is largely dependent on the dimension of the projection space. It has been demonstrated for the Anderson impurity model that using a larger basis can systematically reduce the truncation error~\cite{Fan2018, Fan2019}.
We therefore expect that our results can be improved by enlarging the bases set. In this paper, we focus on a qualitative discussion and the Lacroix's basis is large enough for this purpose.

\subsection{Ferromagnetic case: $R_{\text{pol}} \neq 0$}
\begin{figure}[t!]
	\begin{center}
		\includegraphics[width=1.0\columnwidth]{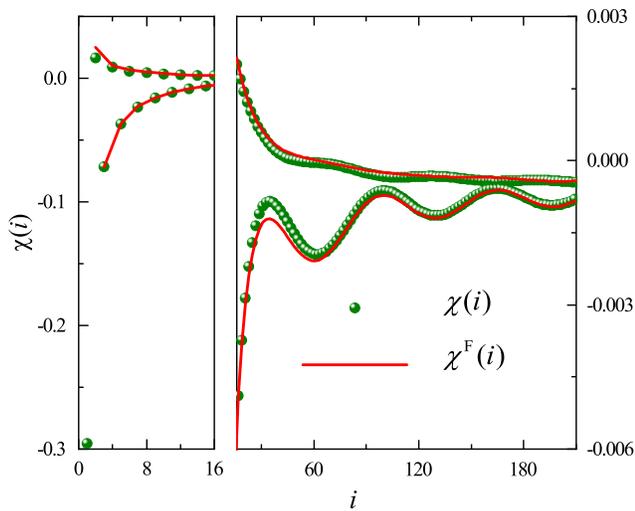}
	\end{center}
	\caption{ (Color online) The equal-time spatial SSCF $\chi(i)$ as a function of distance $i$. The spheres indicate the PTA-EOMGF result, Eq. (\ref{chi}) ; The curves denote the fitting result, Eq. (\ref{chi_f}). We use $N=709$, $T=0.0$, $U=2.0$, $\varepsilon_{d} = -U/2$, $V = 0.85$ and $h = 0.2$. The ratio of the conduction electrons been polarized is $R_{\text{pol}}\approx-0.03$. Because $\chi(i)$ reduces rapidly with $i$, we show the result in two panels. The left (right) panel shows the result in the short-range (long-range) part. } \label{Fig4}
\end{figure}

\begin{figure}[t!]
	\begin{center}
		\includegraphics[width=1\columnwidth]{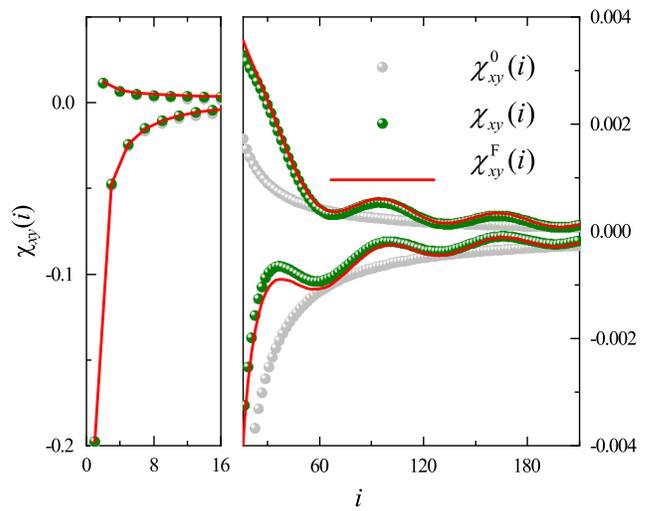}
	\end{center}
	\caption{ (Color online) The dependence of $\chi_{xy}(i)$ on distance $i$. The olive (light-gray) spheres indicate the PTA-EOMGF result, Eq. (\ref{chixy_k_to_i}) for the case $R_{\text{pol}}\neq 0$ ($R_{\text{pol}}= 0$); The curves denote the fitting result, Eq. (\ref{fitchixy}). We use the same parameters as those given in the caption of Fig. \ref{Fig4}. For clarity, the short-range (long-range) part of the result is shown in the left (right) panel. } \label{Fig5}
\end{figure}

\begin{figure}[t!]
	\begin{center}
		\includegraphics[width=1\columnwidth]{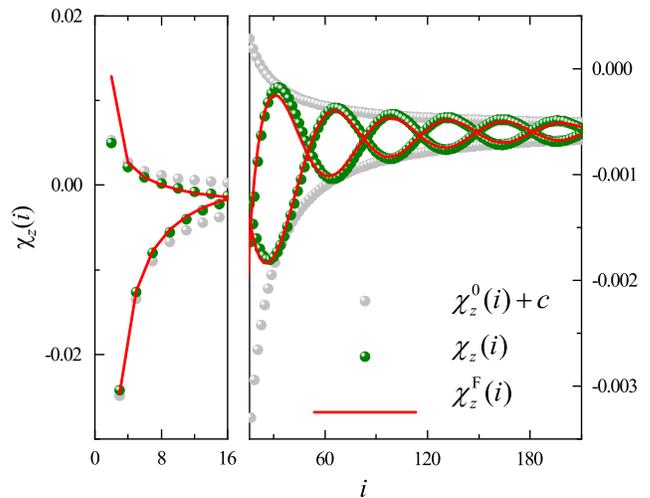}
	\end{center}
	\caption{ (Color online) The dependence of $\chi_{z}(i)$ on distance $i$. The olive (light-gray) spheres indicate the PTA-EOMGF result, Eq. (\ref{chiz_k_to_i}) for the case $R_{\text{pol}}\neq 0$ ($R_{\text{pol}}= 0$); The curves denote the fitting result, Eq. (\ref{fitchiz}). To compare the results ($R_{\text{pol}}\neq 0$ and $R_{\text{pol}}= 0$), we move $\chi^{0}_{z}(i)$ by $c=-5.39\times 10^{-4}$. We use the same parameters as those given in the caption of Fig. \ref{Fig4}. For clarity, the short-range (long-range) part of the result is shown in the left (right) panel.  } \label{Fig6}
\end{figure}

\begin{figure}[t!]
	\begin{center}
		\includegraphics[width=1\columnwidth]{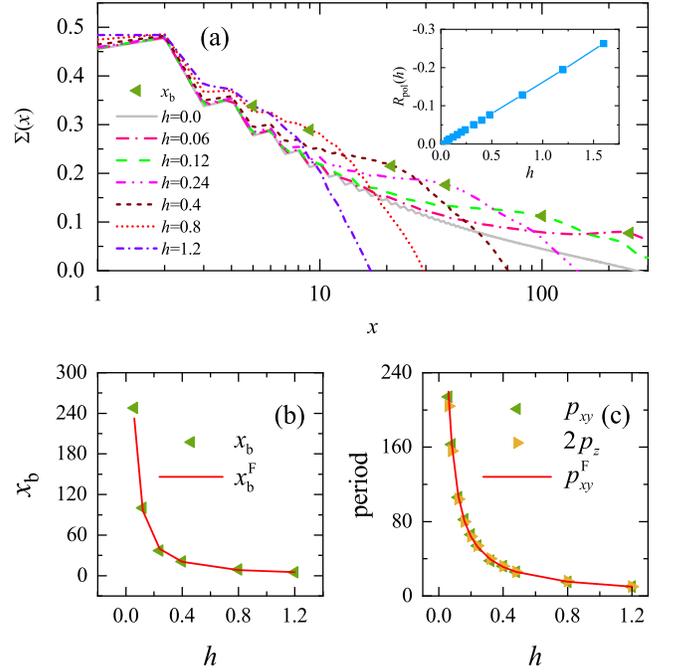}
	\end{center}
	\caption{ (Color online) (a) $\Sigma(x)$ as a function of $x$ for various $h$ values. The scatters show the position $x_{\text{b}}$ of the  bump, which appears for all of the curves $h\neq 0$.   The inset shows the dependence of $R_{\text{pol}}(h)$ on $h$. (b) The dependence of $x_{\text{b}}$ on $h$. The red curve shows the fitting result of $x^{\text{F}}_{\text{b}} = 6.36h^{-1.28}$. (c) The dependence of $p_{xy}$ and $2p_{z}$ on $h$. The red curve shows the fitting result of $p^{\text{F}}_{xy}=12.18h^{-1.03}$. We take $T=0.0$, $U=1.0$, $\varepsilon_{d} = -U/2$ and $V = 0.7$.} \label{Fig7}
\end{figure}

For the ferromagnetic case, an effective magnetic field $h$ is applied to the conduction electrons.
The spin of the conduction electrons is partially polarized.
In Fig. \ref{Fig4},  we apply $h = 0.2$ and obtain a conduction electrons polarization $R_{\text{pol}}\approx-0.03$ (see the inset of Fig. \ref{Fig7}(a)). The minus sign means that the spin is anti-parallel to the external field. The equal-time spatial SSCF $\chi(i)$ at $T=0$ as a function of position $i$ is calculated and shown in Fig. \ref{Fig4} (spheres).
It is found that in the short range regime, $\chi(i)$ behaves similarly to that of the paramagnetic case shown in the inset of Fig. \ref{Fig2}(b). Besides the even-odd oscillation, a wave-like pattern arises in the long-range regime due to bath spin polarization.
To get an insight into the pattern in the long distance, two components of $\chi(i)$, $\chi_{xy}(i)$ and $\chi_{z}(i)$, are calculated and shown in Figs. \ref{Fig5} and \ref{Fig6}, respectively.
Shown together in Figs. \ref{Fig5} and \ref{Fig6} are the zero-field curves and our fitting curves (to be discussed below).
Due to breaking of SU(2) symmetry by the effective field, $\chi_{xy}(i)$ and $\chi_{z}(i)$ behave quite differently in the long-range regime.
In the short-range regime ($i \lesssim 16$ in Figs. \ref{Fig5} and \ref{Fig6}), both $\chi_{xy}(i)$ and $\chi_{z}(i)$ are close to their zero-field counterpart, $\chi^{0}_{xy}(i)$ and $\chi^{0}_{z}(i)$, respectively. In the long-range regime ($i \gtrsim 16$), the envelope of $\chi_{xy}(i)$ has a periodic wave-like pattern, while that of $\chi_{z}(i)$ shows a wave-packet pattern. $\chi_{xy}(i)$ seems to obtain an oscillating envelope component superimposed on $\chi^{0}_{xy}(i)$. The magnitude of this component decays with increasing $i$. In contrast, the wave-packet pattern in $\chi_{z}(i)$ can only be described as an oscillating factor multiplied to the smoothly decaying envelope of $\chi^{0}_{z}(i)$, giving rise to additional nodes in the envelope curve.

These features in $\chi_{xy}(i)$ and $\chi_{z}(i)$ can be traced back to the roles of bath spin polarization and Kondo screening. In both Figs. \ref{Fig5} and \ref{Fig6}, comparison with zero-field quantities reveals that the oscillating envelope patterns solely arise from the spin polarization, i.e. the splitting of the bath Fermi surface.
The periods $p_{xy}$ and $p_{z}$ of the envelope oscillation in $\chi_{xy}(i)$ and $\chi_{z}(i)$ are found to scale as $p_{xy} = 2 p_{z} \sim 1/h$ (see Fig. \ref{Fig7}(c) and the discussion in Appendix B). It is also observed that the envelopes of $\chi_{xy}(i)$ and $\chi_{z}(i)$ are both confined by the corresponding zero-field curves. This indicates that the decay of the magnitude of the envelop oscillation is governed mainly by the Kondo effect. As a check to this conclusion, we have calculated $\chi(i)$, $\chi_{xy}(i)$ and $\chi_{z}(i)$ at $U=0$, where the Kondo effect is absent. We find that the long-range oscillating pattern remains, but the magnitude is one order of magnitude smaller than that at $U \neq 0$ where Kondo effect is present.

Many theories have been developed to explore the spatial SSCF, including perturbative method, renormalization group (RG) method, and conformal field theory method. As for perturbative method, it is only applicable at high temperature. The RG method focuses much on the short-range correlations~\cite{Chen1992}. Barzykin and Affleck have developed a renormalization group improved perturbative technique to calculate the spatial SSCF, but it still cannot access the region $T<T_{\text{K}}$. The conformal field theory approach succeeds in dealing with the low temperature, long distance spin-spin correlation, but failed in dealing with the correlation in the Kondo screening region~\cite{Barzykin1998}. To get some insight into the low temperature, short distance spin-spin correlation, Affleck and others considered the multichannel Kondo model~\cite{ Affleck1991, Ludwig1994, Affleck1993}. It is difficult in theory to find an exact formula in the region $T<T_{\text{K}}$, which can describe $\chi(i)$ as a function of distance $i$ in both the short- and long-range regions. Our results shown in Figs. \ref{Fig5} and \ref{Fig6} provide a unique opportunity to draw some quantitative knowledge for the SSCF in low temperatures. Below, we intend to extract a formula that is able to fit the numerical result of $\chi(i)$ in both the short- and long-range regions. The roles of  the Kondo screening and bath spin polarization in SSCF are encoded into the formula. The fitting results for $\chi^{\text{F}}(i)$, $\chi^{\text{F}}_{xy}(i)$, and $\chi^{\text{F}}_{z}(i)$ are shown as red curves in Figs. \ref{Fig4}, \ref{Fig5}, and \ref{Fig6}, respectively.

Due to breaking of SU(2) symmetry, we seek for two formulas $\chi_{xy}^{\text{F}}(i)$ and $\chi_{z}^{\text{F}}(i)$ that can fit $\chi_{xy}(i)$ and $\chi_{z}(i)$, respectively. $\chi^{\text{F}}(i)$ is then obtained by
\begin{equation} \label{chi_f}
\chi^{\text{F}}(i) = \chi^{\text{F}}_{xy}(i) + \chi^{\text{F}}_{z}(i),\,\,\,\,\ (i\ge2).
\end{equation}
Comparing $\chi^{0}_{xy}(i)$ with $\chi_{xy}(i)$ in Fig. \ref{Fig5}, we find that the local minimums of $\chi_{xy}(i)$ coincide with $\chi^{0}_{xy}(i)$ and their difference is a cosine function with a decaying amplitude. We therefore propose the following formula,
\begin{equation} \label{fitchixy}
\chi^{\text{F}}_{xy}(i) = \chi^{0}_{xy}(i) +  \Upsilon(i).
\end{equation}
The function $\Upsilon(i)$ due to spin polarization reads
\begin{equation}
\Upsilon(i) =f(i) \left[\cos\left(  \frac{2\pi}{p_{xy}} i \right)  + 1 \right],
\end{equation}
where $p_{xy}$ is the period of the envelope oscillation. For best fitting of the curve, we found that the decaying function $f(i)$ must be a power function
\begin{equation} \label{fi}
f\left( i \right) = c_{1}i^{\alpha} + c_{2}.
\end{equation}
In the long-range limit, $\lim_{i\to \infty}\Upsilon(i) =c_{2} \cos\left(  2\pi i / p_{xy} \right)  + c_{2}$. Hence, $c_{2}$ is the wave amplitude at the long-range limit. In Fig. \ref{Fig5}, we fit the numerical result of $\chi_{xy}(i)$ with Eq. (\ref{fitchixy}), resulting in $\alpha = -1.5$, $c_{1} = 0.12 $, $c_{2} = 5.9\times 10^{-5}$ and $p_{xy} = 66$. Note that $c_2$ is negligibly small.

In Fig. \ref{Fig6}, $\chi^{0}_{z}(i)$ is moved downward by $c$ to compare with $\chi_{z}(i)$. We find that the two branches of envelope of $\chi^{0}_{z}(i)+c$ contains that of $\chi_{z}(i)$ in the middle. The spin polarization does not result in any additional decay function in the fitting formula of $\chi_{z}(i)$. Considering an oscillating factor multiplied to $\chi^{0}_z(i)$ and taking care of the difference on even and odd sites (the upper envelope of $\chi^{0}_z(i)$ is composed of $i$ even, and the lower one of $i$ odd), we arrive at the following fitting formula for $\chi_{z}(i)$,
\begin{equation} \label{fitchiz}
\chi^{\text{F}}_{z}(i) = \chi^{1}_{z}(i)  + \chi^{2}_{z}(i) + \kappa(i),\,\,\,\,\ (i\ge2)
\end{equation}
where
\begin{align}
\chi^{1}_{z}(i) &= \beta \chi^{0}_{z}(i)  \cos\left( \frac{\pi}{p_{z}} i \right), \label{fitchiz1} \\
\chi^{2}_{z}(i) &= \gamma \chi^{0}_{z}(i-1)  \cos\left[ \frac{\pi}{p_{z}}\left( i - 1\right) \right] \cos\left[\pi\left(i -1 \right)\right], \label{fitchiz2}  \\
\kappa(i) &= \frac{1}{4} \left( 2\chi^{0}_{z}(i) + \chi^{0}_{z}(i+1) + \chi^{0}_{z}(i-1) \right) + c.  \label{fitchizk}
\end{align}
Here, $\kappa(i)$ is the average hight of the curve $\chi^{0}_{z}(i)$ within two sites interval plus a tiny shift $c$. It accurately describes the hight of the $\chi_{z}(i)$ curve at node points.
$\chi^{1}_{z}(i)$ and $\chi^{2}_{z}(i)$ are introduced to handle the up-down asymmetry of the envelope function of $\chi^{0}_{z}(i)$. For details of the fitting for $\chi_z(i)$, see Appendix B.
In this equation, the fitting parameters are $\beta$ and $\gamma$. $p_{z}$ is the period of the wave-packet.  In Fig. \ref{Fig6}, we use Eq. (\ref{fitchiz}) to fit $\chi_{z}(i)$ and obtain $\beta = 0.7$, $\gamma = 0.3$, $c=-5.39\times 10^{-4}$ and $p_{z} = 33$.

Figure \ref{Fig4} shows the fitting result of $\chi(i)$. The main features induced by spin partial polarization are captured by the fitting formulas. The fitting curves for $\chi^{\text{F}}_{xy}(i)$ and $\chi^{\text{F}}_{z}(i)$ are shown in Figs. \ref{Fig5} and \ref{Fig6}, respectively. They agree quite well with the corresponding numerical results. The biggest deviation between $\chi(i)$ and $\chi^{\text{F}}(i)$ occurs at the neighbor of $i=30$. Because we take $i\ge2$ in Eq. (\ref{fitchiz}), $\chi(i)$ and $\chi_{z}(i)$ at $i=1$ have not been fitted by the corresponding fitting formulas.These fitting results show that at the present accuracy level, the Kondo screening mainly determines the decay of SSCF through $\chi_{z}^{0}(i)$, while the bath spin polarization produces the oscillation in the envelope by an additive or multiplicative terms for $\chi_{xy}(i)$ and $\chi_{z}(i)$, respectively.

We studied the dependence of all the fitting parameter on $U$ and $V$ (see Table I in Appendix C). We find that among all the fitting parameters in Eqs. (\ref{fitchixy})-(\ref{fi}) and Eqs. (\ref{fitchiz})-(\ref{fitchizk}), $\alpha \approx -1.5$ is independent on $U$ and $V$ within the allowed fitting error range, which indicates that the decay rate of the envelope oscillation in $\chi_{xy}(i)$ is universal.
Our present finding that  $-2<\alpha<-1$ suggests that the spin polarization makes the crossover weaker, considering that $\chi^{0}(i)$ crosses over from $i^{-1}$ to $i^{-2}$ behaviors around $\xi_{\text{K}}$~\cite{Borda2007, Barzykin1998}. Therefore, it would be more difficult to determine the boundary of the Kondo screening region when the spin of the conduction electrons is partially polarized.
When we scan $U$ and $V$, we also find that both $p_{xy}$ and $p_{z}$ are independent on $U$ and $V$, which indicates that the envelope oscillation only results from the splitting of the Fermi surface.
In Figure \ref{Fig7}(c), $p_{xy}$ and $2p_{z}$ are shown as functions of $h$. We find $p_{xy} = 2p_{z}$, as expected (see Appendix B). The result of $p_{xy}$ is fitted by $p^{\text{F}}_{xy} = 12.18h^{-1.03}$, indicating $p_{xy} \sim 1/h$. At the limit $h=0$, the period diverges. Hence, in paramagnetic case the period pattern is absent.

The above fitting formula, Eqs.(\ref{fitchixy}) and (\ref{fitchiz}), and the $1/h$ dependence of the oscillation period of the envelope function have a nice interpretation from the beating of two Friedel oscillations associated to two spin-split Fermi surfaces of conduction electrons. To show this, we consider approximate solution of $\chi_{xy}(i)$ [Eq.(\ref{chixy})] and $\chi_{z}(i)$ [Eq.(\ref{chiz})] at $U=0$. The finite $h$ splits the Fermi surfaces of spin up and down conduction electrons. Using Wick's theorem and borrowing the formula for a quadratic bath electron dispersion\cite{Mezei1972},  we obtain $\langle d_{\sigma}^{\dagger} a_{i\sigma} \rangle \propto \cos(i k_{F\sigma}a)$. It describes the Friedel oscillation due to the existence of Fermi surface, with the period modulated by the Fermi momentum $k_{F\sigma}$ of spin-$\sigma$ conduction electrons. $k_{F\sigma}$ is determined by equation
\begin{equation}  \label{Friedel1}
     -2t \cos(k_{F\sigma}a) = \mu - \frac{1}{2}\sigma h.
\end{equation}
Using $\mu=0$ and expanding $k_{F\sigma}$ at $h=0$, we obtain $k_{F\sigma} = k_F - \sigma h /(4 t a)$, with $k_F=\pi/(2a)$.
Finally, we obtain
\begin{eqnarray}  \label{Friedel2}
    \chi_{xy}(i) & \propto & \cos(i k_{F\uparrow} a) \cos(i k_{F \downarrow} a)  \nonumber \\
                 &=& \cos(2k_{F}a i) + \cos\left(\frac{h}{2t}i \right).
\end{eqnarray}
For $\chi_z(i)$, we obtain
\begin{eqnarray}   \label{Friedel3}
    \chi_{z}(i) & \propto &  \cos^{2}(i k_{F\uparrow} a) + \cos^{2}(i k_{F \downarrow} a)  \nonumber \\
                 &=& \cos(2k_{F}a i) \cos\left(\frac{h}{2t}i \right).
\end{eqnarray}

These results have strong similarity to our fitting formula Eqs.(\ref{fitchixy}) and (\ref{fitchiz}): additive correction in $\chi^{F}_{xy}$ and multiplicative correction in $\chi^{F}_z(i)$. The fitted period shown in Fig.\ref{Fig7}(c), $p^{F}_{xy}=2p^{F}_{z}=12.18 h^{-1.03}$, agrees well with the result $p_{xy} = 2p_{z} = 4\pi/h = 12.57/h$ from the above analysis.
We therefore get strong support to our fitting formula, and confirm that the additional oscillation in the envelope function of SSCF arises from the beating of two Friedel oscillations. The Kondo screening effect is mainly embodied in the spatial decay behavior and the coefficient of the oscillation. As to be discussed below, the influence of Zeeman field on Kondo screening is more easily observed in the spatial distribution of the integrated SSCF  $\Sigma(x)$.

Figure \ref{Fig7}(a) shows the field dependence of $\Sigma(x)$ on $h$. For a fixed $h$, $\Sigma(x)|_{h\neq 0}$ follows the curve of $\Sigma(x)|_{h=0}$ in the small $x$ range and deviates from $\Sigma(x)|_{h=0}$ when $x$ is larger than a crossover scale $x_{\text{b}}$, at which a huge bump appears. For $x \gg x_b$, $\Sigma(x)$ decreases linearly with increasing $x$. For sufficiently large $x$, $\Sigma(x) <0$ occurs (not shown). Apparently, the bump and the subsequent linear decrease in $\Sigma(x)|_{h \neq 0}$ originates from the bath spin polarization induced by $h$. When the bath Fermi surface is split by the field $h$, the population of spin up and down electrons on the impurity site is redistributed to find the lowest energy for the whole system. This induces a net impurity spin anti-parallel to those of the bath and gives rise to a negative $\chi_{z}(i)$. In the large $x$ limit where the bath electrons are mainly those located at Fermi surface, $\chi_{z}(i)$ tends to a negative constant in large $i$, leading to a linearly decreasing $\Sigma(x)$.

The position of the bump $x_b$ moves to the impurity with increasing $h$. When $h$ tends to zero, the bump moves to infinity and $\Sigma(x)|_{h=0}$ is recovered.
In Fig. \ref{Fig7}(a), the positions $x_{\text{b}}$ of the bump for all of the curves are marked by triangles. We show $x_{\text{b}}$ as a function of $h$ and its fitting result $x^{\text{F}}_{\text{b}}= 6.36h^{-1.28}$ in Fig. \ref{Fig7}(b).

\begin{figure}[t!]
\vspace{-3.0cm}
	\begin{center}
		\includegraphics[width=1.3\columnwidth]{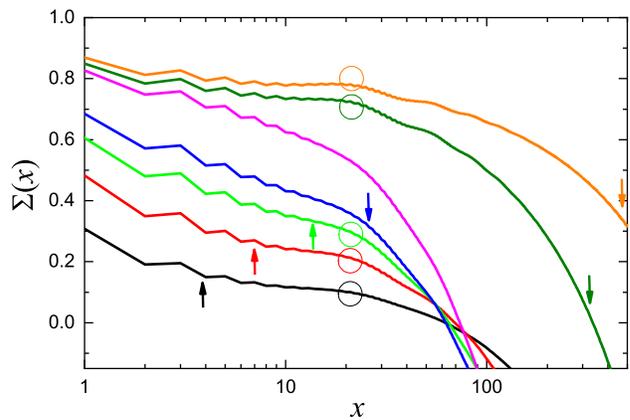}
	\end{center}
    \vspace{-0.5cm}
	\caption{ (Color online) $\Sigma(x)$ as a function of $x$ for various $U/\Gamma$ values at fixed $h=0.4$. From bottom to up on the left, $U/\Gamma= 0.510$, $2.041$, $3.781$, $5.355$, $10.796$, $11.755$, and $12.755$. We use fixed $U\Gamma = 0.49$ and $\varepsilon_d=-U/2$. The arrorws mark out the position of $\xi_\text{K}(h=0)$ estimated using $\xi_\text{K} = p/T_\text{K}$, with $p=2.2$ and $T_\text{K}$ from Eq. (\ref{xi_k}). The positions of bumps at $x_b$ are marked out by circles.  } \label{Fig8}
\end{figure}

It is tempting to relate the bump position $x_{\text{b}}$ to the size of the Kondo cloud for system with a spin polarized bath. However, this is not the case at least for small $h$. At $h=0$, $x_b = \infty$ while $\xi_K$ recovers its paramagnetic value $\xi_K^0 \equiv \hbar v_F/T_K$ and it is finite. To investigate this issue, we investigate in Fig.8 the evolution of $\Sigma(x)$ with $U/\Gamma$, for fixed $h=0.4$ and $U\Gamma=0.49$. According to Eq.(\ref{xi_k}), the paramagnetic Kondo cloud size $\xi_K^{0}$ increases exponentially with increasing $U/\Gamma$. In Fig. \ref{Fig8}, we mark out the estimated $\xi_\text{K}^{0}$ by an arrow for each $U/\Gamma$ value. The circles mark out the bump that separates the small $x$ regime where $\Sigma(x)$ is identical to $h=0$ case and the large $x$ regime where $\Sigma(x)$ decreases linearly. We find that $x_{\text{b}}$ does not depend on $U/\Gamma$, while $\xi_K^0$ increases exponentially with $U/\Gamma$. At $\xi_K^{0} \sim x_b$, $\Sigma(x)$ crossover smoothly from small $x$ to large $x$ regimes, and the bump is absent (see $U/\Gamma = 5.355$ and $10.796$ in Fig.8).

A scenario of the competition between Kondo screening and bath polarization can be drawn as the following. As stated above, the linear decrease of $\Sigma(x)$ is a signal of spin polarization of bath electrons near Fermi surface. For $h \gg T_{K}$, this amounts to breaking of Kondo screening between impurity electron and the bath electrons, leading to $x_{\text{b}} \ll \xi_K^0$. In this case, the Kondo screening cloud is suppressed and the actual Kondo screening length is dominated by $x_{\text{b}}$. On the other hand, for $h \ll T_{K}$, the spin polarization of bath electrons near Fermi surface is not sufficient to break the screening of impurity spin by the high energy part of bath electrons. In this case, $x_{\text{b}} \gg \xi_K^0$ and the actual Kondo screening length is still dominated by $\xi_K^0$. Therefore, we conclude that for $h \gg T_K$, the bump position at $x_{\text{b}}$ marks out the real space boundary between the Kondo screened regime ($x < x_{\text{b}}$) and the regime where Kondo screening is broken by bath spin polarization. While for $h \ll T_K$, the $x_{\text{b}}$ only reflect the real space boundary between spin polarized and unpolarized bath electrons.

In Fig.\ref{Fig7}(a), we obtain $x_{\text{b}} \sim h^{-1.28}$, with an exponent close to $-1$. Considering that our data for small $h$ is less accurate due to finite size effect, we believe that the present result is in line with the following picture. In the absence of $h$, $\Sigma(x)$ has a decay range of $\xi_{\text{K}}^0 \sim 1/T_{\text{K}}$ which controls the Kondo screening length. For $h \ll T_{\text{K}}$, the polarization does not influence the Kondo screening length. While for $h \gg T_{\text{K}}$, the size of the Kondo screening cloud is suppressed and dominated by $h$, leading to the expectation that $\xi_{\text{K}} \sim \min(1/T_\text{K}, 1/h)$.

\section{Summary}
We extended the formulation of PTA-EOMGF into real space to calculate the equal-time spatial SSCF of impurity and bath electrons for Anderson impurity model with a spin polarized bath. The results are benchmarked with DMRG data at zero bath bias $h=0$ and qualitative agreement is reached. For finite $h$, due to spin polarization of bath electrons close to Fermi energy, an oscillation emerges in the envelope function of SSCF curve, with period proportional to $1/h$. The corresponding integrated SSCF curve shows a bump at $x_{\text{b}} \sim h^{-1.28}$ which marks the boundary between Kondo screened regime and the Kondo-broken regime. Based on our data, we propose fitting formula for the transverse and longitudinal SSCF, respectively. They perfectly fit $\chi_{xy}(i)$ and $\chi_{z}(i)$ in both the short- and long-range regions. A simple description of the influence of Kondo screening and bath spin polarization on SSCF is thus obtained.

\section{Acknowledgements}
This work is supported by NSFC (Grant No. 11974420). This work is supported by NSFC (Grant Nos. 11974348, and 11834014) and the Fundamental Research Funds for the Central Universities. Z.G.Z. is supported in part by the National Key R\&D Program of China (Grant No. 2018FYA0305800), the Strategic Priority Research Program of CAS (Grant Nos. XDB28000000, and XDB33000000), and the Major Research plan of the National Science Foundation of China (Grant No. 92165105).

\appendix{}
\setcounter{figure}{0}
\renewcommand{\thefigure}{S\arabic{figure}}
\section{ SSCFs: $\chi_{xy}(i)$ and $\chi_{z}(i)$ }
In this section, we give formula for $\chi_{xy}(i)$ and $\chi_{z}(i)$. According to the definition of the equal-time spatial SSCF, we have
\begin{equation} \label{A1}
\chi(i) = \langle s^{x}_{i} s^{x}_{d} \rangle + \langle s^{y}_{i} s^{y}_{d} \rangle + \langle s^{z}_{i} s^{z}_{d} \rangle.
\end{equation}
Here, the direction of the effective magnetic field $h$ is in $z$-direction. Because of $h$, the SU(2) symmetry of spin is broken in $z$ direction. In $xy$-plane, the spin is still isotropic.
In terms of $s^{\pm}_{i} = s^{x}_{i} \pm i s^{y}_{i}$, we have
\begin{equation} \label{A2}
\chi(i) = \frac{1}{2}\left( \langle s^{+}_{i} s^{-}_{d} \rangle + \langle s^{-}_{i} s^{+}_{d} \rangle \right) + \langle s^{z}_{i} s^{z}_{d} \rangle.
\end{equation}
It is obvious that $\langle s^{+}_{i} s^{-}_{d} \rangle = \langle s^{-}_{i} s^{+}_{d} \rangle$, because $s^{+}_{i} s^{-}_{\text{d}} = \left( s^{-}_{i} s^{+}_{\text{d}} \right)^{\dagger}$ and the average $\langle \cdots \rangle$ is real. Therefore,
\begin{equation} \label{A3}
\chi(i) = \chi_{xy}(i) + \chi_{z}(i).
\end{equation}
Here,
\begin{equation} \label{A4}
\chi_{xy}(i) = \langle s^{+}_{i} s^{-}_{d} \rangle
\end{equation}
denotes the transverse spin-spin correlation and
\begin{equation} \label{A5}
\chi_{z}(i) = \langle s^{z}_{i} s^{z}_{d} \rangle
\end{equation}
denotes the longitudinal spin-spin correlation. The $\alpha$ $(\alpha = x,\,y,\,z)$ component of $\vec{s}_{i}$ is
\begin{equation} \label{A6}
s^{\alpha}_{i} = \frac{1}{2} \sum_{\sigma \sigma^{\prime}} a^{\dagger}_{i\sigma} \bm{\sigma}^{\alpha}_{\sigma\sigma^{\prime}} a_{i\sigma^{\prime}},
\end{equation}
where $\bm{\sigma}^{\alpha}$ is the Pauli matrices.
Hence, we have
\begin{equation} \label{A7}
s^{+}_{i} = a^{\dagger}_{i \uparrow} a_{i \downarrow},\,\,\,\ s^{-}_{i} = a^{\dagger}_{i \downarrow} a_{i \uparrow},\,\,\,\ s^{z}_{i} = \frac{1}{2} \left( n_{i\uparrow} - n_{i\downarrow} \right).
\end{equation}
In the same way, the spin operators of the impurity are
\begin{equation} \label{A8}
s^{+}_{\text{d}} = d^{\dagger}_{ \uparrow} d_{ \downarrow},\,\,\ s^{-}_{\text{d}} = d^{\dagger}_{ \downarrow} d_{ \uparrow},\,\,\ s^{z}_{\text{d}} = \frac{1}{2} \left( n_{\text{d}\uparrow} - n_{\text{d}\downarrow} \right).
\end{equation}
Substituting Eq. (\ref{A7}) and Eq. (\ref{A8}) to Eq. (\ref{A4}) and Eq. (\ref{A5}), we obtain
\begin{equation}
\chi_{xy}(i) = \langle a^{\dagger}_{i\uparrow} a_{i\downarrow} d^{\dagger}_{\downarrow} d_{\uparrow} \rangle,
\end{equation}
and
\begin{equation}
\chi_{z}(i) =
\frac{1}{4} \sum_{\sigma} \left(\langle n_{i\sigma} n_{d\sigma} \rangle - \langle n_{i\sigma}n_{d\bar{\sigma}} \rangle\right).
\end{equation}

\section{Numerical results: $\kappa(i)$, $\chi^{1}_{z}(i)$ and $\chi^{2}_{z}(i)$}
\begin{figure}[t!]
	\begin{center}
		\includegraphics[width=1\columnwidth]{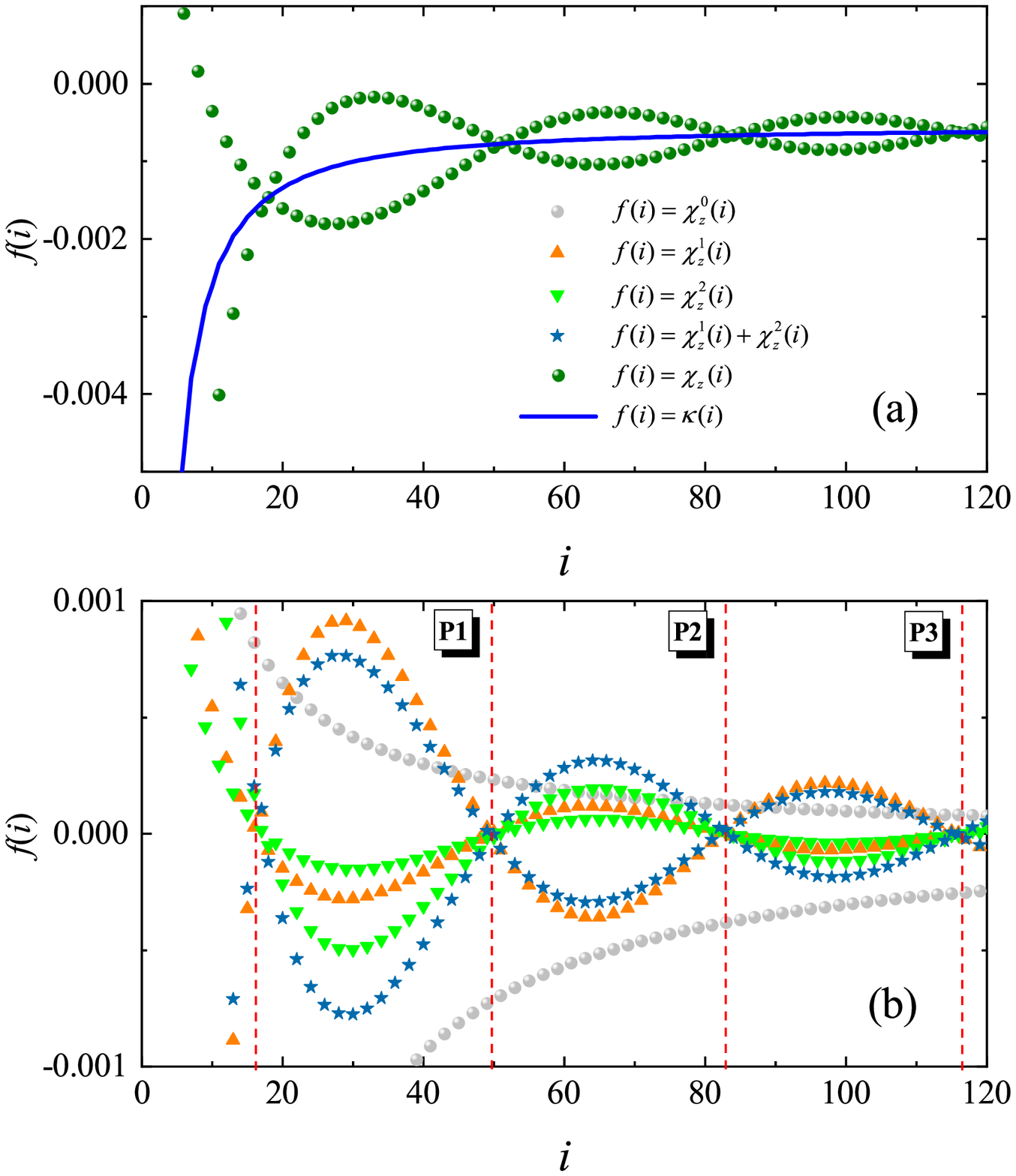}
	\end{center}
	\caption{(Color online) The dependence of $f(i)$ on the distance $i$. (a) shows the numerical results of $\chi_{z}(i)$ and $\kappa(i)$. (b) shows the numerical results of $\chi^{0}_{z}(i)$, $\chi^{1}_{z}(i)$, $\chi^{2}_{z}(i)$ and $\chi^{1}_{z}(i)+\chi^{2}_{z}(i)$. This is to illustrate the contribution of $\chi^{1}_{z}(i)$ and $\chi^{2}_{z}(i)$ to fitting the numerical result of $\chi_{z}(i)$. We show three periods (labeled by P1, P2, P3), which are divided by the red dash lines. We use the same parameters as shown in the caption of Fig. \ref{Fig4}.}\label{S1}
\end{figure}
In Fig. \ref{S1}, we show the numerical results of $\chi^{1}_{z}(i)$, $\chi^{2}_{z}(i)$ and $\kappa(i)$, which are the components of $\chi^{\text{F}}_{z}(i)$, defined by Eqs. (\ref{fitchiz1})-(\ref{fitchizk}) respectively. To conveniently discuss the individual contribution to the fitting of $\chi_{z}(i)$ data, we also show $\chi_{z}(i)$, $\chi^{0}_{z}(i)$ and $\chi^{1}_{z}(i) + \chi^{2}_{z}(i)$.
In Fig. \ref{S1}(a), we see that $\kappa(i)$ perfectly fits the node points of $\chi_{z}(i)$. The curve $\chi_{z}(i)-\kappa(i)$ is almost symmetric about zero. We use $\chi^{0}_{z}(i)$ multiplied with cosine functions to fit it. $\chi^{0}_z(i)$ has an even-odd oscillation. The upper and lower envelopes are composed of data points at even and odd $i$, respectively. They are not symmetric about zero. This makes it necessary to fit $\chi_{z}(i)-\kappa(i)$ with two cosine functions multiplied with each envelope of $\chi^{0}_z(i)$. The one-site shift in the formula of $\chi^{1}_{z}(i)$ and $\chi^{2}_{z}(i)$ are for this purpose. The additional factor $\cos[\pi(i-1)]$ in the formula of $\chi^{2}_{z}(i)$ is to turn the two terms into same phase. For $i$ even, $\chi^{1}_{z}(i)$ uses the upper envelope of $\chi^{0}_{z}(i)$ and $\chi^{2}_{z}(i)$ uses the lower one, and vice versa for $i$ odd.

In Fig. \ref{S1}(b), three periods (P1, P2 and P3) are shown.
For $\chi_z(i)$ fit, as shown in Eqs. (\ref{fitchiz1})-(\ref{fitchiz2}), the cosine function changes sign when $i$ increases by $p_z$ sites. This leads to interchange of upper and lower envelope curves and forms an effective period of $p_z$ in the fitting function $f(i)$. The two adjacent peaks of the fitting function are sitting on an even and an odd sites, respectively. Note that in the fitting formula for $\chi_{xy}(i)$ Eq.(\ref{fitchixy}), the period is $p_{xy}$. This explains the fact that out fitting always gives $p_{xy}=2 p_z$ (see Fig.\ref{Fig7}(c)).
The total contribution of $\chi^{1}_{z}(i)$ and $\chi^{2}_{z}(i)$ is thus
\begin{align} \label{B1}
f(i) &= \chi^{1}_{z}(i) + \chi^{2}_{z}(i) \nonumber \\
       &= \beta \chi^{0}_{z}(i)  \cos\left( \frac{\pi}{p_{z}} i \right)  + \gamma \chi^{0}_{z}(i-1)  \nonumber \\
      &\,\,\,\,\ \times \cos\left[ \frac{\pi}{p_{z}}\left( i - 1\right) \right] \cos\left[\pi\left(i -1 \right)\right].
\end{align}

\vspace{1cm}

\section{Fitting results}
In Table \ref{table1}, we list the value of the parameters in the fitting formulas.
We fit $\chi_{xy}(i)$ with the first $211$ data points. As for $\chi_{z}(i)$, we fit with the first $111$ data points.
\begin{table*}[btp]
\renewcommand\arraystretch{2}
\centering
\caption{The value of parameters in the fitting formulas} \label{table1}
\begin{tabular*}{18cm}{@{\extracolsep{\fill}}lcccccccccc}
\hline\hline
 U        & V     & $\alpha$  & $c_{1}$ & $c_{2}$ & $p_{xy}$ & $\beta$ & $\gamma$ & $c$ & $p_{z}$  \\
\hline
    0.5  & 0.7  & -1.6 &  0.11  & $3.07\times 10^{-5}$ & 66  & 0.62 & 0.4 & $-3.16\times 10^{-4}$ & 31 \\
\hline
    1.0  & 0.7  & -1.4 &  0.08 & $3.43\times 10^{-5}$ &  66  & 0.66 & 0.4 & $-4.68\times 10^{-4}$ & 32 \\
\hline
   1.5  & 0.7  & -1.5  &  0.14  & $5.20\times 10^{-5}$ & 65  & 0.68 & 0.38 & $-6.67\times 10^{-4}$ & 33 \\
\hline
  2.0  & 0.7  & -1.6 &  0.22  & $1.40\times 10^{-4}$ & 66 &  0.7 &   0.38 &    $-9.00\times 10^{-4}$ & 33 \\
  \hline
  2.5  & 0.7  & -1.6 &  0.27  & $1.90\times 10^{-4}$ & 66 &  0.72 & 0.36 & $-1.19\times 10^{-3} $ & 33\\
  \hline
  3.0  & 0.7  & -1.5&  0.23& $2.03\times 10^{-4}$ & 66 & 0.72 & 0.34  & $-1.42\times 10^{-3} $ & 34 \\
  \hline
  3.5  & 0.7  & -1.6&  0.33 & $2.20\times 10^{-4}$ & 67  & 0.72 & 0.32 & $-1.52\times 10^{-3} $ & 35 \\
  \hline
  2.0  & 0.5  & -1.5 &  0.25 & $2.42\times 10^{-4}$ & 67  & 0.74 & 0.38 & $-2.27\times 10^{-3} $ & 34 \\
  \hline
  2.0  & 0.55  & -1.7 &  0.41 & $2.70\times 10^{-4}$ & 66  & 0.72 & 0.36 & $-1.89\times 10^{-3} $ & 34 \\
  \hline
  2.0  & 0.6  & -1.7 &  0.38 & $2.48\times 10^{-4}$ & 66  & 0.7 & 0.36 & $-1.49\times 10^{-3} $ & 33 \\
  \hline
  2.0  & 0.65  & -1.7 &  0.33 & $1.82\times 10^{-4}$ & 67  & 0.7 & 0.36 & $-1.16\times 10^{-3} $ & 33\\
   \hline
  2.0  & 0.75  & -1.5 &  0.16 & $6.20\times 10^{-5}$ & 65 & 0.7 & 0.38 & $-7.57\times 10^{-4} $ & 33\\
   \hline
  2.0  & 0.8  & -1.6 &  0.18 & $7.40\times 10^{-5}$ & 66 & 0.7 & 0.38 & $-6.5\times 10^{-4} $ & 33 \\
   \hline
  2.0  & 0.85  & -1.5 &  0.12 & $5.90\times 10^{-5}$ & 66  & 0.7 & 0.38 & $-5.39\times 10^{-4} $ & 33 \\
\hline\hline
\end{tabular*}
\end{table*}
%

%

\end{document}